\documentclass[aps,prd,twocolumn,showpacs,groupedaddress,nofootinbib,,superscriptaddress]{revtex4-1}
\usepackage{graphicx}
\usepackage{amssymb,amsmath,pifont}

\begin{document}


\title{No stable dissipative phantom scenario in the framework of a complete cosmological dynamics}




\author{Norman Cruz}
\email[]{norman.cruz@usach.cl}
\affiliation{Departamento de F\'{\i}sica, Universidad de Santiago de Chile, Casilla 307, Santiago, Chile}

\author{Samuel Lepe}
\email[]{slepe@ucv.cl}
\affiliation{Instituto de F\'isica, Pontificia Universidad Cat\'olica de Valpara\'iso\\
Casilla 4950, Valpara\'iso, Chile}

\author{Yoelsy Leyva}
\email[]{yoelsy.leyva@uta.cl}
\affiliation{Departamento de F\'isica, Facultad de Ciencias, Universidad de Tarapac\'a, Arica, Chile}
\affiliation{Instituto de F\'isica, Pontificia Universidad Cat\'olica de Valpara\'iso\\
Casilla 4950, Valpara\'iso, Chile}

\author{Francisco Pe\~na}
\email[]{francisco.pena@ufrontera.cl}
\affiliation{Departamento de Ciencias F\'isicas. Facultad de Ingenier\'ia y Ciencias, Universidad de la Frontera\\
Casilla 54-D, Temuco, Chile}

\author{Joel Saavedra}
\email[]{joel.saavedra@ucv.cl}
\affiliation{Instituto de F\'isica, Pontificia Universidad Cat\'olica de Valpara\'iso\\
Casilla 4950, Valpara\'iso, Chile}

\date{\today}

\begin{abstract}

We investigate the phase space dynamics of a bulk viscosity model in the Eckart approach for a spatially flat
Friedmann-Robertson-Walker universe. We have included two barotropic fluids and a dark energy component.
One of the barotropic fluids is treated as an imperfect fluid having bulk viscosity, whereas the other components are assumed to behave as perfect fluids. Both barotropic fluids are identified as either radiation or dark matter. Considering that the bulk viscosity acts on either radiation or dark matter, we find that viscous phantom solutions with stable behavior are not allowed in the framework of complete cosmological dynamics. Only an almost zero value of the bulk viscosity allows a transition from a radiation-dominated to a matter-dominated epoch, which then evolves to an accelerated late time expansion, dominated by dark energy. 
\end{abstract}

\pacs{98.80.-k, 95.35.+d, 95.36.+x, 98.80.Jk}
\maketitle


\section{Introduction}
Observational evidence indicates that the present acceleration of the Universe may be successfully explained by a cosmic fluid with negative pressure, a concept which has been baptized dark energy. This exotic component contributes about $68 \%$ \cite{Ade2013} of the total energy of the Universe.  Evidence for dark energy is provided by several complementary probes such as the high redshift surveys of supernovae \cite{Riess2011}, the cosmic microwave background (CMB) \cite{Komatsu2011, Hinshaw2013, Ade2013}, and the integrated Sachs-Wolfe effect \cite{Giannantonio2008, Giannantonio2012}.

Dark energy is considered a fluid characterized by a negative
pressure and is usually represented by the equation of state $w\approx-1$.\footnote{$w=p/
\rho$, where $p$ and $\rho$ are the pressure and energy density of the fluid.} This is confirmed by the latest Planck results \cite{Ade2013}, which give
$w = -1.13^{+0.24}_{-0.25}$ and $w = -1.09\pm0.17$ ($2\sigma$ C.L.) by using
CMB data\footnote{These CMB data are referred to as the combination of the WMAP polarization low multipole likelihood \cite{Bennett2013} + Planck temperature power spectrum \cite{Ade2013a}.} combined with 
BAO \cite{Percival2010, Padmanabhan2012, Blake2011, Anderson2013, Beutler2011} and Union2.1~\cite{Suzuki2012} data, respectively, for
a constant $w$ model. However, when the same CMB data are combined with SNLS compilation\footnote{A sample of $473$ type Ia supernovae.} \cite{Conley2011} and the Hubble constant $H_{0}$ \cite{Riess2011a},
the equation of state (EOS) for this dark component are $w = -1.13^{+0.13}_{
-0.14}$ and $w = -1.24^{+0.18}_{-0.19}$ ($2\sigma$ C.L.), respectively. 

The phantom behavior, 
derived from dynamical scalar fields, has been studied in~\cite{Wang2012, Novosyadlyj2012}. In \cite{Wang2012}, the free parameters of phantom scalar field models with power law and exponential potentials were constrained using data compiled from
CMB \cite{Komatsu2011}, BAO \cite{Cole2005, Percival2010}, and Union2.1 \cite{Suzuki2012}. These results, when considered 
together with those obtained in a similar analysis for quintessence models \cite{Wang2012a}, and using the goodness of fit and information criteria, allowed the authors to conclude that the cosmological constant is more preferable, from the statistical point of view, than phantom and quintessence models, although the phantom dark energy model is slightly better at fitting the observational data than the quintessence. However, in \cite{Novosyadlyj2012}, the best-fit values of the parameters for dark energy models, in which the phantom scalar field initially mimics a cosmological constant term to finally and slowly evolve toward the big bip singularity, are determined jointly with all other cosmological parameters by the Markov chain Monte Carlo method \cite{Lewis2002} using observational data on cosmic microwave background anisotropies and polarization \cite{Jarosik2011, Larson2011}, Ia type supernovae \cite{Guy2007, Kessler2009, Sullivan2011}, BAO \cite{Percival2010, Suzuki2012}, big bang nucleosynthesis \cite{Steigman2007, Wright2007}, and Hubble constant measurements from HST \cite{Riess2009}.
Similar computations have been carried out for $\Lambda$CDM and quintessence scalar field models of dark energy \cite{Novosyadlyj2012}. It has been shown that the current data slightly prefer the phantom model, but the differences in the maximum likelihoods are not statistically significant. As such, the possibility of phantom behavior for the dark energy fluid cannot be discarded.


Furthermore, an important result, and prior to the discovery of the present speed-up of the Universe, was the fact that a dissipative mechanism like bulk viscosity may give rise to an accelerated evolution of the Universe \cite{heller, Barrow1986a, Diosi1984, Waga1986, Padmanabhan1987, Gron1990, Maartens1995a, Zimdahl1996a}. At the same time, bulk viscosity provides the only dissipative mechanism consistent, in a homogeneous and isotropic background, with the cosmological principle. Thus, it has been proposed as one of the possible ways to induce an accelerated phase in the evolution of the Universe, namely, early time inflation \cite{Campo2010a, Bastero-Gil2012, Setare2013e, Setare2014b} or the present accelerated period \cite{Cataldo2005, Brevik2005, Li2009,Avelino2009, Avelino2010, Velten2011, Velten2012, Velten2013a, Velten2013, Avelino2013}. 

In a  Friedmann-Robertson-Walker (FRW) universe, the inclusion of bulk viscosity allows the possibility of violating the dominant energy condition \cite{Barrow1986a, Barrow1988c} and hence phantom solutions. In the Eckart approach \cite{Eckart1940}, bulk viscosity introduces dissipation by only redefining the effective pressure of the cosmic fluid, namely,
\begin{eqnarray}\label{effecti}
P_{_{eff}}= p+\Pi= p-3 \zeta H,
\end{eqnarray}
where $p$ is the kinetic pressure of the cosmic fluid, $\Pi$ is the bulk viscous pressure, $H$ is the Hubble parameter, and the bulk viscosity coefficient, $\zeta$, satisfies\footnote{For a fluid on a FRW geometry, the local entropy production is defined as $T\nabla_{\mu}s^{\mu}=9H^{2}\zeta$ \cite{misner1973gravitation}, where $\nabla_{\mu}s^{\mu}$ is the rate of entropy production in a unit volume and $T$ is the temperature of the fluid. Since the second law of thermodynamics provides that $\nabla_{\mu}s^{\mu}\geq0$, then for an expanding Universe ($H>0$) $\zeta\geq0$.}
\begin{equation}
 \zeta\geq0,\label{LSLT};
\end{equation}
this latter requirement guarantees nonviolation of the local second law of thermodynamics (LSLT) \cite{Maartens1996, Zimdahl2000}.

Since the equation of energy balance states that
\begin{eqnarray}\label{ConsEq}
\dot{\rho}+ 3 H (\rho+p+\Pi)=0,
\end{eqnarray}
the violation of dominant energy condition, i.e., $\rho+p+\Pi<0$, implies an increasing energy density of the fluid that fills the Universe.

Phantom behavior due to the presence of a bulk viscosity has been investigated in many cosmological contexts: in~\cite{Cataldo2005}, a big rip singularity solution was obtained; the full causal Israel-Stewart-Hiscock framework theory \cite{Israel1979} assumed a late time universe filled with only one barotropic fluid. Furthermore, the late time evolution of a Chaplygin gas model with bulk viscosity, in a causal and truncated version of the Israel-Stewart formalism and in the Eckart approach, was studied in \cite{Cruz2007}. In both frameworks, the authors found new types of future singularities. However, a viable phantom solution was derived only in the truncated approach. In the context of the Eckart approach, the following investigations have been addressed:  big rip singularities for various forms of $w = w(\rho)$ and the bulk viscosity $\zeta = \zeta(\rho)$~\cite{Brevik2013a}; little rip cosmologies~\cite{Brevik2011a, Gorbunova2010a}; phantom crossing in modified gravity~\cite{Brevik2005, Brevik2005a}; and unified dark fluid cosmologies ~\cite{Li2013a, Pourhassan2013a, Saadat2013, Saadat2013a, Setare2010a}. In addition, the conditions for the physical viability of a cosmological model in which dark matter has bulk viscosity and also interacts with dark energy were discussed in \cite{Avelino2013}. In this case, the model took into account radiation, baryons, dark matter and dark energy and considered a general interaction term between the dark components. In relation to the viscosity, the authors assumed the ansatz $\zeta\propto\zeta_{0}H$, where $H$ is the Hubble parameter and $\zeta_{0}$ a positive constant. Joint analysis of the phase space of the model and the observational test shows that (a) complete cosmological dynamics requires either null or negative bulk viscosity; (b) observations consistently point to a negative value of the bulk viscous coefficient; (c) the phantom nature of dark energy ($w_{de}<-1$) was consistently suggested by the cosmological observations. The first two results found in \cite{Avelino2013} rule out any viscous model with the ansatz $\zeta\propto\zeta_{0}H$, despite the latter result being in line with recent observations \cite{Ade2013}.

More recently, a deep discussion concerning degeneracy in bulk viscosity models when $\zeta\equiv\zeta(H)$ was presented in \cite{Velten2013}. \footnote{The model discussed in \cite{Avelino2013} is an specific case of $\zeta\equiv\zeta(H)$.} In addition, the cases where $\zeta_{i}$ of some fluid $i$ (e.g. radiation or dark matter) depends on its own energy density $\zeta_{i}\equiv\zeta_{i}(\rho_{i})$ were discussed. The existence of phantom solutions for this latter Ansatz was studied using multiple observational tests. However, as a consequence of the degeneracy criterion used, it was not possible to obtain, in the observational analysis performed by the authors, the best-fit values/signs of parameters of the proposed viscosity models,\footnote{Constraints on the dark matter viscosity can be found in \cite{Velten2012} for two models: a) a constant $\zeta$, and b) $\zeta\propto\zeta_{0}/\sqrt{\rho_{m}}$. } namely eg. the coefficient of the bulk viscosity.

In the present work we are interested in studying the phase space of the models proposed in \cite{Velten2013}. Our main goal is to explore if the presence of a phantom solution is compatible with the so-called complete cosmological dynamics \cite{Avelino2013, Leon2014}. According to this framework, at early enough times, all physically viable models must allow the existence of radiation-  and  matter-dominated periods previous to the present accelerated expansion of the Universe.  The phantom solution is obtained by adding bulk viscosity (as a new extra imperfect pressure) to either radiation or pressureless matter \cite{Velten2013}. 

The organization of the paper is as follows: in Sec. II we present the field equations for a flat FRW universe filled with two barotropic fluids and dark energy obeying a barotropic EOS.  We assume that one of the barotropic fluids presents bulk viscosity, and the effective pressure is treated within the framework of the Eckart theory \cite{Eckart1940}. The bulk viscous coefficient is taken to be proportional to the square root of the energy density of fluid taken with an imperfect pressure. In Sec. III, we analyze the evolution equations from the perspective of dynamical systems. We studied two models by matching the barotropic fluids with radiation and dark matter. A detailed scheme of the critical points and their conditions upon the parameters of the model is shown. Finally, Sec. IV is devoted to conclusions.

\section{The Model}\label{SectionDynamicalSystems}

We study a cosmological model in a spatially flat FRW metric, in which the matter components are
two barotropic fluids and dark energy (DE). One of the barotropic fluids and the DE are assumed as perfect fluids, whereas the remaining fluid is treated as an imperfect fluid having bulk viscosity.

The Friedmann constraint and the conservation equations for the matter fluids can be written as
\begin{subequations}
\label{eq:1} 
  \begin{eqnarray}
    H^2 &=& \frac{8 \pi G}{3}\left( \rho_{\rm 1}  + \rho_{\rm 2} + \rho_{\rm de} \right) \, , \label{ConstrainFriedmann} \\
    \dot{\rho}_{\rm 1} &=& - 3 \gamma_{1} H\rho_{\rm 1}+9H^2 \zeta \, , \label{ConsEqRadiation} \\
    \dot{\rho}_{\rm 2}  &=& - 3\gamma_{2}H \rho_{\rm 2}   \, , \label{EqConservationEffective4} \\
    \dot{\rho}_{\rm de} &=& - 3H\gamma_{\rm de}\rho_{\rm de},  \label{EqConservationEffective4a}
  \end{eqnarray}
\end{subequations}  
where $G$ is the Newton gravitational constant; $H$ the Hubble parameter;
$(\rho_{\rm 1}$, $\rho_{\rm 2}$, $\rho_{\rm de})$ are the energy
densities of barotropic and DE fluid components, respectively; and 
$\gamma_{\rm de}$ is the barotropic index of the equation of state (EOS)
of DE, which is defined from the relationship $p_{\rm de} = (\gamma_{\rm de} -1)
\rho_{\rm de}$, where  $p_{\rm de}$ is the pressure of DE. 
The term $9H^2\zeta$ in Eq.~(\ref{ConsEqRadiation}) corresponds to the
bulk viscous pressure of the first barotropic fluid, with $\zeta$ the bulk
viscous coefficient. The nature of the barotropic fluids and the evolution of the model are determined once the value of barotropic indices ($\gamma_{1}$, $\gamma_{2}$) are set; eg., $\gamma=1$ corresponds to pressureless matter (dark matter) and $\gamma=4/3$ represents radiation. 
 
We take the bulk viscous coefficient $\zeta$ to be proportional to the energy density of the first barotropic fluid in the form
\begin{equation}
  \label{ViscosityDefinition}
  \zeta = \zeta_{0}\left(\frac{\rho_{1}}{\rho_{1 0}}\right)^{\alpha},
\end{equation}
where $\zeta_0$ and $\alpha$ are dimensionless constants and $\rho_{10}$ is the present day value of the energy density of the first barotropic fluid. The choices  $\alpha=0$ and $\alpha=-1/2$ have been studied in the literature \cite{Velten2011, Velten2012, Velten2013a}. Both values allow for the reduction of the problems of the integrated Sachs-Wolfe effect for viscous UDM\footnote{Unified dark matter models; see \cite{Velten2012} and references therein.} models \cite{Velten2011}. However, from a dynamical system point of view, it is possible to study the dynamics of the field equations ~(\ref{eq:1})
and~(\ref{eq:R}) for arbitrary values of $\alpha$. It is possible to cover the whole $\alpha$ parameter space with two cases, namely, $\alpha=1/2$ and $\alpha\neq1/2$. The first choice leads to the simplest mathematical problem with a two-dimensional phase space (see Sec. \ref{sec:dynam-syst-persp} ), while for $\alpha\neq1/2$ we obtain a more complex three-dimensional dynamical system. For mathematical simplicity, henceforth we will only study the case $\alpha=1/2$. \footnote{We will address the case $\alpha\neq 1/2$ in a forthcoming paper.}
Finally, the Raychaudhuri equation of the model is
\begin{equation}
  \dot{H} = -4\pi G \left( \gamma_{1} \rho_{\rm 1}  +
\gamma_{2} \rho_{\rm 2} + \gamma_{\rm de} \rho_{\rm de} -3H\zeta \right) \, .
\label{eq:R}
\end{equation}

\section{The dynamical system   perspective  \label{sec:dynam-syst-persp}}
In order to study all possible cosmological scenarios of the model, we
proceed to a dynamical system analysis of Eqs.~(\ref{eq:1})
and~(\ref{eq:R}). Let us first define the set of dimensionless
variables
\begin{subequations}
\label{var}
\begin{equation}
  x=\Omega_{de}\equiv\frac{8\pi G}{3H^{2}} \rho_{\rm de} \, , \quad y = \Omega_{1}\equiv\frac{8\pi G}{3H^{2}} \rho_{\rm 1}. 
 \end{equation}
\end{subequations}
Then, the equations of motion can be written in the following,
equivalent, form:
\begin{subequations}
\label{ev}
\begin{eqnarray}
  \frac{dx}{dN} &=& x \left(3 y \left(\gamma _1-\gamma _2\right)-3 (-1+x) \left(\gamma _2-\gamma _{\text{de}}\right)\right)-{} \nonumber \\
  & & {} x\sqrt{y} \xi, \label{evx}\\
  \frac{dy}{dN} &=& 3y\left((-1+y)\left(\gamma _1-\gamma _2\right)-x\left(\gamma _2-\gamma _{\text{de}}\right)\right)-{}\nonumber\\
  & & {} (-1+y)\sqrt{y}\xi, \label{evy}
\end{eqnarray}
\end{subequations}
where the derivatives are with respect to the $e$-folding number $N
\equiv \ln a$ and
\begin{equation}
\xi=\frac{24\pi G }{H_{0}\Omega_{10}^{\frac{1}{2}}}\zeta_{0},
\end{equation}
where the condition (\ref{LSLT}) implies [see Eq. (\ref{ViscosityDefinition})]
\begin{equation}
 \xi\geq0.\label{2da}
\end{equation}

In term of the new variables, the Friedmann
constraint~(\ref{ConstrainFriedmann}) can be written as:
\begin{equation}  
  \Omega_{\rm 2} = \frac{8\pi G}{3H^2} \rho_{\rm 2} = 1-x-y \, , \label{FCC}
\end{equation}
and then we can choose $(x,y)$ as the only independent dynamical
variables. 

Taking into account that $0\leq\Omega_{\rm 2}\leq 1$,\footnote{Recall that, once we set the values of $\gamma_{1}$ and $\gamma_{2}$ we will able to identify $\Omega_{1}$ and $\Omega_{2}$ with the dimensionless density parameters of radiation and dark matter , if we set $\gamma_{1}=4/3$ and $\gamma_{2}=1$, and vice versa, if we set $\gamma_{1}=1$ and $\gamma_{2}=4/3$.} and imposing the
conditions that both the barotropic fluids and DE components be positive, definite, and  bounded at all times, we can define the phase space of
Eqs.~(\ref{ev}) as
\begin{eqnarray}
  \Psi &=& \{(x,y): 0 \leq 1-x-y \leq 1 ,  0\leq x \leq1 \, ,
  \nonumber \\
  && 0 \leq y \leq 1\} \, . \label{eq:space}
\end{eqnarray}
Other cosmological parameters of interest are the total effective EOS, 
$w_{eff}$, and the deceleration parameter, $q = - (1+\dot{H}/H^2)$,
which can be written, respectively, as
\begin{subequations}
\label{eq:2}
\begin{eqnarray}
  w_{eff} &=& -1-\frac{\sqrt{y} \xi }{3}+y\left(\gamma _1-\gamma _2\right)+\gamma _2-{}\nonumber\\
  & & {} x\left(\gamma _2-\gamma _{\text{de}}\right) , \label{effecw} \\
  q &=& \frac{1}{2} \left(-2-\sqrt{y} \xi +3 y \left(\gamma _1-\gamma _2\right)+3\gamma _2-{}\right.\nonumber\\
  & & {}\left. 3x\left(\gamma _2-\gamma _{\text{de}}\right)\right) \,
  . \label{decel}
\end{eqnarray}
\end{subequations}

\subsection{The case with $\gamma_{1}=1$ (dark matter) and $\gamma_{2}=4/3$ (radiation)}
This case corresponds to a scenario where dark matter (DM) is treated as an imperfect fluid having bulk viscosity with a null hydrodynamical pressure. At the same time, it is clear from the definition of the dimensionless variables that $x=\Omega_{de}$, $y=\Omega_{1}=\Omega_m$ and $\Omega_{2}=\Omega_r$. The full set of critical points of the autonomous system (\ref{evx})-(\ref{evy}), the existence conditions, and the stability conditions for $\gamma_{1}=1$ and $\gamma_{2}=4/3$ are summarized in Table \ref{tab1}. The eigenvalues of the linear perturbation matrix associated to each of the critical points and some important physical parameters are given in Table \ref{tab2}.

\label{CP}
\begin{center}
\begin{table*}[t!]\caption[Qf]{Location, existence conditions
    according to the physical phase space (\ref{eq:space}), and
    stability of the critical points of the autonomous system
    (\ref{evx})-(\ref{evy}) for $\gamma_{1}=1$ and $\gamma_{2}=4/3$. The
    eigenvalues of the linear perturbation matrix associated to each
    of the following critical points are displayed in
    Table~\ref{tab2}.}
  \begin{tabular}{@{\extracolsep{\fill}}| c | c | c | c | c | c |}
    \hline\hline
    $P_i$ & $x$ & $y$ & Existence & Stability \\
    \hline\hline
   $P_{1}$ & $0$ & $0$ & Always & Unstable if $\gamma_{de}<1$ and $ \xi\geq 0$ \\ \hline
	         	 
$P_{2}$ & $0$ & $1$ & Always & Stable if $\gamma _{\text{de}}<1$ and $\xi >3-3 \gamma _{\text{de}}$\\
	          &     &     &        & Saddle if $\gamma _{\text{de}}<1$ and $0\leq\xi <3-3 \gamma _{\text{de}}$ \\\hline
$P_{3}$ & $1-\frac{\xi^{2}}{9(-1+\gamma_{de})^{2}}$ & $\frac{\xi^{2}}{9(-1+\gamma_{de})^{2}}$ & $\gamma _{\text{de}}<1$ and $ 0\leq\xi\leq3(1-\gamma_{de})$ & Stable if $\gamma _{\text{de}}<1$ and $0<\xi <3-3 \gamma _{\text{de}}$ \\
       
	\hline
	
	 $P_{4}$ & $1$ & $0$ & Always & Stable if $\gamma_{de}<1$ and $\xi = 0$\\
	      		&     &     &        & Saddle if $\gamma_{de}<1$and $\xi >0$ \\\hline
   
\end{tabular}\label{tab1}
\end{table*}
\end{center}

\begin{table*}[t!]\caption[Qf]{Eigenvalues and some basic physical
    parameters for the critical points listed in Table~\ref{tab1}, see
    also Eqs.~(\ref{var}) and~(\ref{eq:2}).}
  \begin{tabular}{@{\extracolsep{\fill}}| c | c | c | c | c | c | c |}
    \hline \hline
    $P_i$&$\lambda_1$ & $\lambda_2$ & $\Omega_r$&$w_{eff}$ & $q$\\
    \hline\hline
  $P_{1}$ & $4-3\gamma_{de}$ & sgn($\xi$)$\infty$ & $1$ & $\frac{1}{3}$ & $1$ \\ \hline
  $P_{2}$ & $-1-\xi$ & $3-3\gamma_{de}-\xi$ & $0$ & $-\frac{\xi}{3}$ & $\frac{1}{2}-\frac{\xi}{2}$ \\ \hline
  $P_{3}$ & $-4+3 \gamma _{\text{de}}$ & $\frac{1}{6} \left(-9-\frac{\xi ^2}{-1+\gamma _{\text{de}}}+9 \gamma _{\text{de}}\right)$ & $0$ & $-1+\gamma _{\text{de}}$ & $-1+\frac{3 \gamma _{\text{de}}}{2}$\\ \hline
  $P_{4}$ & $-4+3\gamma_{de}$ & sgn($\xi$)$\infty$ & $0$ & $-1+\gamma_{de}$ & $-1+\frac{3\gamma_{de}}{2}$ \\
	
    \hline \hline
\end{tabular}\label{tab2}
\end{table*}


\subsubsection{Critical points and stability \label{sec:crit-points-stab}}
Critical point $P_{1}$ corresponds to a pure radiation-domination era, $\Omega_{r}=1$, and always exists independently of the value$/$sign of the viscosity parameter $\xi$. It also represents a decelerating expansion solution with $w_{eff}=1/3$ and $q=1$. The stability of this critical point is the following:
\begin{itemize}
\item Unstable if $\gamma_{de}<1$ and $\xi\geq0$.
\end{itemize}

$P_{2}$ is a critical point dominated by the pressureless matter component, $\Omega_{m}=1$, and always exists. This critical point exhibits two different stability behaviors, namely,
\begin{itemize}
\item Stable if $\gamma _{\text{de}}<1$ and $\xi >3-3 \gamma _{\text{de}}$, 
\item Saddle if $\gamma _{\text{de}}<1$ and $0\leq\xi <3-3 \gamma _{\text{de}}$.
\end{itemize}

From Table \ref{tab2}, we notice that if $\xi^2\ll 1$, there is a point that corresponds to a standard matter-domination period ($w_{eff}=0$). An interesting fact of $P_{2}$ is the value of EOS parameter $w_{eff}=-\xi/3$. In the stable region ($\gamma _{\text{de}}<1$ and $\xi>3-3\gamma_{de}$), this critical point represents an accelerating phantom solution with $w_{eff}<-1$  if
\begin{equation}
 \xi >3\;\;\;\text{and}\;\;\;\frac{3-\xi }{3}<\gamma _{\text{de}}<1.\nonumber
\end{equation}

Despite this result being in correspondence with those observations that tend to mildly favor a present day value of $w_{eff}$ in the phantom region, it will be shown in the next subsection that it is not possible to associate this behavior with a realistic late time solution.

$P_{3}$ corresponds to a scaling solution between pressureless matter and dark energy and exists when 
\begin{equation*}
\gamma _{\text{de}}<1\;\;\;\text{and}\;\;\; 0\leq\xi\leq3(1-\gamma_{de}).
\end{equation*}
It represents an accelerated solution ($q<0$) if
\begin{equation}
 \gamma_{de}<\frac{2}{3},
\end{equation}
 and shows a stable behavior given that
 \begin{itemize}
  \item $\gamma _{\text{de}}<1$ and $0<\xi <3-3 \gamma _{\text{de}}$.
  \end{itemize}
In the particular case where $\xi\ll1$, a strictly dark energy domination is recovered ($x=\Omega_{de}=1$).

Finally, critical point $P_{4}$ represents a pure dark-energy-dominated solution ($\Omega_{de}=1$) and always exists. The stability of $P_{4}$ is the following,
\begin{itemize}
\item Stable if $\gamma_{de}<1$ and $ \xi = 0$,
\item Saddle if $\gamma_{de}<1$ and $\xi >0$.
\end{itemize}
For a dark energy fluid  with $\gamma_{de}<1$ and $\xi>0$, this critical point is always an accelerated solution with a saddle behavior (see Table \ref{tab2} for further details).

\subsubsection{Cosmology evolution from critical points \label{sec:cosm-evol-from}}

According to the complete cosmological dynamics paradigm (see \cite{Avelino2013, Leon2014} for recent discussion about this topic), our model should describe (a) a radiation-dominated era at early times (RDE), (b) a matter-domination era (MDE) at intermediate stages of the evolution, and finally, (c) the present stage of accelerated expansion of the Universe. The required dominance stages can be translated into critical points, and the desired transitions between them into the heteroclinic orbits that connect two critical points \cite{Heinzle2009, Urena-Lopez2012}.

According to the complete cosmological dynamics basis, one of the critical points of the model should correspond to a RDE, and this point should be unstable in nature. The unstable behavior of this critical point ensures that it can be the source of any orbits in the phase space. The only candidate in our model is the critical point $P_{1}$. In fact, $P_{1}$ satisfies the condition for pure radiation dominance ($\Omega_{r}=1$) and always exists. Its unstable nature, for all realistic dark energy fluids ($\gamma_{de}<1$), is clear in the previous section.  

As we mentioned before, the Universe requires the existence of a matter-dominated era in order to explain the formation of the cosmic structure. This MDE is recovered by a $P_{2}$ ($\Omega_{m}=1$), which always exists. However, as Table \ref{tab2} shows, we cannot recover a standard pressureless matter-dominated picture ($w_{eff}=0$, $q=1/2$) unless $\xi=0$ \footnote{Recall that the second law of thermodynamics requires a nonnegative value of $\xi$; see (\ref{2da}).}. For a non-null value of $\xi$, $P_{2}$ represents a decelerating solution if $0<\xi<1$ ($q>0$) or an accelerating expansion solution if $\xi>1$ ($q<0$); see Table \ref{tab2} for further details. In both regions where $\xi\neq0$, the presence of bulk viscosity leads to an undesirable expansion rate of the Universe and makes it impossible to associate this critical point with a realistic MDE. 

\begin{figure}[h!]
\begin{center} 
\includegraphics[width=21cm]{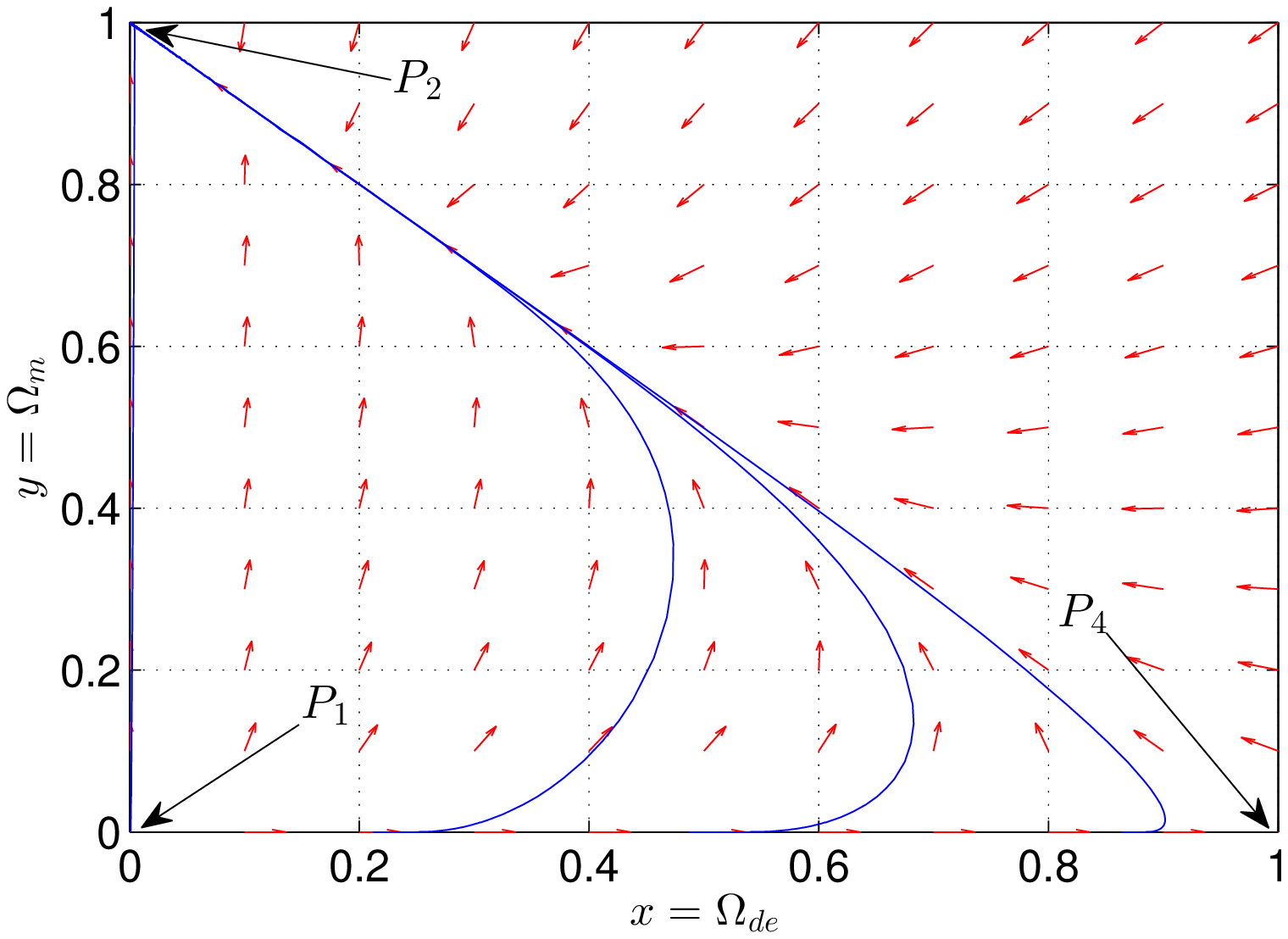} 
\caption{\label{fig4} Vector field in the plane ($x$, $y$) for the autonomous system (\ref{evx})-(\ref{evy}). The free parameters have been chosen to be ($\xi$, $\gamma_{de}$)$=$($3.01$, $0.01$). In this case, the dark-matter-dominated solution, $P_{2}$, is the late time attractor of the system, representing a viscous phantom solution ($w_{eff}=-1.003$). This latter solution restricts the critical points of the system to three and none of them corresponds to a true MDE.}
\end{center}
\end{figure}

An interesting characteristic of $P_{2}$ is that, for certain regions in the parameter space ($\xi$, $\gamma_{de}$), it represents not only an accelerated solution but also a phantom one. As Table \ref{tab2} shows, the presence of this phantom solution ($w_{eff}=-\xi/3<-1$) in $P_{2}$ depends solely on the value of the viscosity parameter $\xi$ and not on the nature of the dark energy component. \footnote{For $P_{2}$ the value of the effective EOS parameter, $w_{eff}=-\xi/3$, is insensitive of the barotropic index of dark energy ($\gamma_{de}$).} However, as Table \ref{tab1} and Table \ref{tab2} show, its stability behavior depends on the value of the barotropic index. \footnote{Recall that the viscous phantom solution of $P_{2}$ exists and is stable if: $\xi >3$ and $(3-\xi)/3<\gamma _{\text{de}}<1$.} This stable and viscous phantom behavior of $P_{2}$ could be important time behavior of the Universe and, at the same time, could be in line with those cosmological observations that moderately favor the present  date EOS  parameter  value in the phantom region $w_{eff}<-1$. However, it is not possible to select initial conditions that connect the RDE ($P_{1}$) to the phantom attractor solution ($P_{2}$) through a true MDE. As we mentioned before, only $P_{2}$ (with $\xi=0$) could play this role. Furthermore, the conditions for the existence and stability of the viscous phantom solution restrict the autonomous system (\ref{evx})-(\ref{evy}) to three critical points, namely $P_{1}$, $P_{2}$ and $P_{4}$, and none of them corresponds to a true MDE. In addition, another strong argument against the existence of this viscous phantom solution represented by $P_{2}$ comes from structure formation. The existence of dark energy and $\xi>3$ do not lead to the correct growth of galaxies, as was shown in \cite{Velten2012, Velten2013a}. Thus, it is not possible to associate this viscous phantom behavior in $P_{2}$ with a realistic late time  solution.  Figure \ref{fig4} shows some example orbits in the plane ($x$, $y$) to illustrate this situation.

\begin{figure}[h!]
\begin{center} 
\includegraphics[width=8cm]{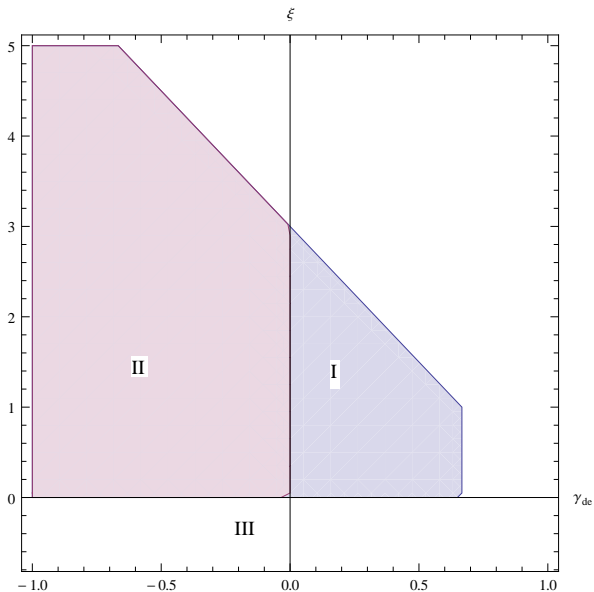}
\caption{\label{fig1} The accelerated and stable region for critical point $P_{3}$ in terms of the parameters ($\xi$, $\gamma_{de}$) is represented by region $I$. Region $II$ represents the zone for which the phantom solution exists and is stable. The lower region, labeled $III$, is forbidden by the LSLT ($\xi<0$).}
\end{center}
\end{figure} 

\begin{figure}[h!]
\begin{center} 
\includegraphics[width=21cm]{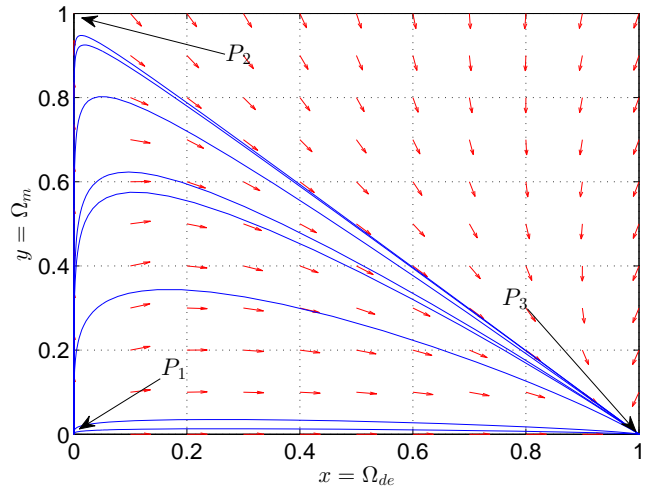}
\caption{\label{fig3}Vector field in the plane ($x$, $y$) for the autonomous system (\ref{evx})-(\ref{evy}). The free parameters have been chosen to be ($\xi$, $\gamma_{de}$)$=$($0.01$, $0.01$). In this case, the DM-DE scaling solution, $P_{3}$, is the late time attractor of the system, representing an accelerated solution. The transition from the RDE ($P_{1}$) to $P_{3}$ allow us to select appropriate initial conditions to recover a true MDE ($P_{2}$) with $w_{eff}\simeq0$ and $q\simeq1/2$.}
\end{center}
\end{figure}

 Another feature of the model is the presence of two more accelerated solutions, described by critical points $P_3$ and $P_{4}$. $P_{3}$ is a scaling solution between dark matter and dark energy and can be an attractor for $\gamma _{\text{de}}<1$ and $0<\xi <3-3 \gamma _{\text{de}}$ whereas $P_{4}$ is always a saddle solution for all realistic dark energy with $\xi>0$. As Fig. \ref{fig1} shows, the existence of an attractor phantom solution in $P_{3}$ depends on the dark energy component; in other words, it is only possible to obtain attractor phantom solutions if the phantom dark energy component, $w_{de}<-1$ ($\gamma_{de}<0$), is present. A favorable scheme would be one in which the initial conditions allow for complete cosmological dynamics, eg., $P_{1}$ $\Longrightarrow$ $P_{2}$ $\Longrightarrow$ $P_{4}$. In terms of the cosmological evolution of the Universe, the above favorable scenario implies that the Universe started at early times from a RDE, then evolved into a MDE, to enter in the final phase of accelerated expansion. Recall that in this favorable scenario, it is necessary to select a sufficiently small value for the bulk viscosity $\xi\ll1$ in order to recover a true MDE ($P_{2}$) and that is not possible to obtain a realistic phantom solution. It is important to emphasize that this finding on the allowed values of $\xi$ corroborates the previous results of \cite{Velten2012, Velten2013a} for the cases of $\alpha=0$ and $\alpha=-1/2$. Figure \ref{fig3} shows some example orbits in the plane ($x$, $y$) to illustrate the above situation.
 
\subsection{The case with $\gamma_{1}=4/3$ (radiation) and $\gamma_{2}=1$ (DM)}
This case corresponds to a scenario where radiation is treated as an imperfect fluid having bulk viscosity. At the same time, it is clear from the definition of the dimensionless variables that $x=\Omega_{de}$, $y=\Omega_{1}=\Omega_r$ and $\Omega_{2}=\Omega_m$. The full set of critical points of the autonomous system (\ref{evx})-(\ref{evy}), and the existence conditions and the stability conditions for $\gamma_{1}=4/3$ and $\gamma_{2}=1$ are summarized in Table \ref{tab3}. The eigenvalues of the linear perturbation matrix associated to each of the critical points and some important physical parameters are given in Table \ref{tab4}.

\begin{table*}[t!]\caption[Qf]{Location, existence conditions
    according to the physical phase space (\ref{eq:space}), and
    stability of the critical points of the autonomous system
    (\ref{evx})-(\ref{evy}) for $\gamma_{1}=4/3$ and $\gamma_{2}=1$. The
    eigenvalues of the linear perturbation matrix associated to each
    of the following critical points are displayed in
    Table~\ref{tab4}.}
  \begin{tabular}{@{\extracolsep{\fill}}| c | c | c | c | c | c |}
    \hline\hline
    $P_i$ & $x$ & $y$ & Existence & Stability \\
    \hline\hline
   $P_{1}$ & $0$ & $1$ & Always & Stable if $\gamma _{\text{de}}<1$ and $\xi >4-3 \gamma _{\text{de}}$ \\ 
    &     &     &        & Unstable if $\gamma _{\text{de}}<1$ and $0\leq \xi <1$ \\
   &     &     &        & Saddle if $\gamma _{\text{de}}<1$ and $1<\xi <4-3 \gamma _{\text{de}}$ \\\hline
           	
$P_{2}$ & $0$ & $\xi^{2}$ & $0< \xi<1$ & Saddle if $\gamma _{\text{de}}<1$ and $0< \xi <1$\\ \hline
$P_{3}$ & $0$ & $0$ & Always & Unstable if $\gamma _{\text{de}}<1$ and $\xi >0$\\
	          &     &     &        & Saddle if $\gamma _{\text{de}}<1$ and $\xi=0$ \\\hline
$P_{4}$ & $1-\frac{\xi ^2}{\left(4-3 \gamma _{\text{de}}\right){}^2}$ & $\frac{\xi ^2}{\left(4-3 \gamma _{\text{de}}\right){}^2}$ & $\gamma _{\text{de}}<1$ and $0< \xi \leq 4-3 \gamma _{\text{de}}$ & Stable if $\gamma _{\text{de}}<1$ and $0<\xi <4-3 \gamma _{\text{de}}$ \\
       
	\hline
 $P_{5}$ & $1$ & $0$ & Always & Stable if $\gamma_{de}<1$ and $\xi = 0$\\
	      		&     &     &        & Saddle if $\gamma_{de}<1$and $\xi >0$ \\\hline
   
\end{tabular}\label{tab3}
\end{table*}

\begin{table*}[t!]\caption[Qf]{Eigenvalues and some basic physical
    parameters for the critical points listed in Table~\ref{tab3}; see also Eqs.~(\ref{var}) and~(\ref{eq:2}). We have used the definition $A=\text{Abs}\left[8+\xi ^2+9 \left(-2+\gamma _{\text{de}}\right) \gamma _{\text{de}}\right]$.}
  \begin{tabular}{@{\extracolsep{\fill}}| c | c | c | c | c | c | c |}
    \hline \hline
    $P_i$&$\lambda_1$ & $\lambda_2$ & $\Omega_m$&$w_{eff}$ & $q$\\
    \hline\hline
  $P_{1}$ & $1-\xi$ & $4-\xi -3 \gamma _{\text{de}}$ & $0$ & $\frac{1}{3}-\frac{\xi }{3}$ & $1-\frac{\xi }{2}$ \\ \hline
  $P_{2}$ & $\frac{1}{2} \left(-1+\xi ^2\right)$ & $3-3 \gamma _{\text{de}}$ & $1-\xi^{2}$ & $0$ & $\frac{1}{2}$ \\ \hline
  $P_{3}$ & sgn($\xi$)$\infty$ & $3-3 \gamma _{\text{de}}$ & $1$ & $0$ & $\frac{1}{2}$ \\ \hline
  $P_{4}$ & $\frac{1}{4} \left(-10+9 \gamma _{\text{de}}+\frac{-\xi ^2+A}{-4+3 \gamma _{\text{de}}}\right)$ & $\frac{1}{4} \left(-10+\frac{\xi ^2+A}{4-3 \gamma _{\text{de}}}+9 \gamma _{\text{de}}\right)$ & $0$ & $-1+\gamma _{\text{de}}$ & $-1+\frac{3 \gamma _{\text{de}}}{2}$\\ \hline
  $P_{5}$ & $-4+3\gamma_{de}$ & sgn($\xi$)$\infty$ & $0$ & $-1+\gamma_{de}$ & $-1+\frac{3\gamma_{de}}{2}$ \\
	
    \hline \hline
\end{tabular}\label{tab4}
\end{table*}


\subsubsection{Critical points and stability}
Critical point $P_{1}$ corresponds to a radiation-dominated era ($\Omega_{r}=1$). The stability of this critical point is the following:
\begin{itemize}
 \item Unstable if $\gamma _{\text{de}}<1$ and $0\leq \xi <1$,
 \item Saddle if $\gamma _{\text{de}}<1$ and $1<\xi <4-3 \gamma _{\text{de}}$,
 \item Stable if $\gamma _{\text{de}}<1$ and $\xi >4-3 \gamma _{\text{de}}$.
 \end{itemize}
As expected, a background level at this point represents a true decelerating RDE if $\xi\ll1$ ($w_{eff}\approx \frac{1}{3}$ and $q\approx1$). If this condition is satisfied, $P_{1}$ behaves as an unstable solution. However, $P_{1}$ also mimics, at background level, dark matter ($w_{eff}=0$) and phantom fluids ($w_{eff}<-1$) if the bulk viscosity is $\xi=1$ and $\xi>4$, respectively. If the phantom solution exists, then $P_{1}$ will behave as a stable late time solution of the autonomous system  (\ref{evx})-(\ref{evy}). We will discuss in more details the cosmological implications of this critical point in the next subsection.

Hyperbolic critical point $P_{2}$ corresponds to a scaling solution between radiation and dark matter and exists if $0<\xi<1$, being a decelerated solution $q=1/2$. From a stability point of view, this critical point always has a saddle behavior demanding that
\begin{itemize}
 \item $\gamma_{de}<1$ and $0<\xi<1$.
\end{itemize}
At the limit, when $\xi\ll1$, this point recovers a true DME ($\Omega_{m}=1$).

Critical point $P_{3}$ corresponds to a pure DM dominance solution ($\Omega_{m}=1$, $w_{eff}=0$) and always exists. A striking aspect of this solution is its dynamical behavior: for a non-null bulk viscosity on the radiation fluid, $P_{2}$ behaves as an unstable solution (see Table \ref{tab3}). This means that it is not possible to associate $P_{3}$ with an intermediate stage of matter domination in the evolution of the Universe. 

The hyperbolic critical point $P_{4}$ represents a scaling solution between radiation and DE. In addition, it is a stable solution for
\begin{itemize}
 \item $\gamma_{de}<1$ and $0<\xi<4-3\gamma_{de}$.
\end{itemize}
As Table \ref{tab4} shows, $P_{4}$ is an accelerated solution if
\begin{equation}\label{ccco}
 -1+\frac{3}{2}\gamma_{de}<0, 
\end{equation}
recovering strictly dark energy domination if $\xi\ll1$.

Finally, $P_{5}$ corresponds to a pure dark energy  domination solution ($\Omega_{de}=1$) and always exists. For a non null bulk viscosity, this critical point represents a saddle solution if 
\begin{itemize}
 \item $\gamma_{de}<1$ and $\xi>0$.
\end{itemize}
Just like the previous critical point, $P_{5}$ represents an accelerated solution if (\ref{ccco}) is satisfied.
 
\subsubsection{Cosmology evolution from critical points}
\begin{figure}[h!]
\begin{center} 
\includegraphics[width=24cm]{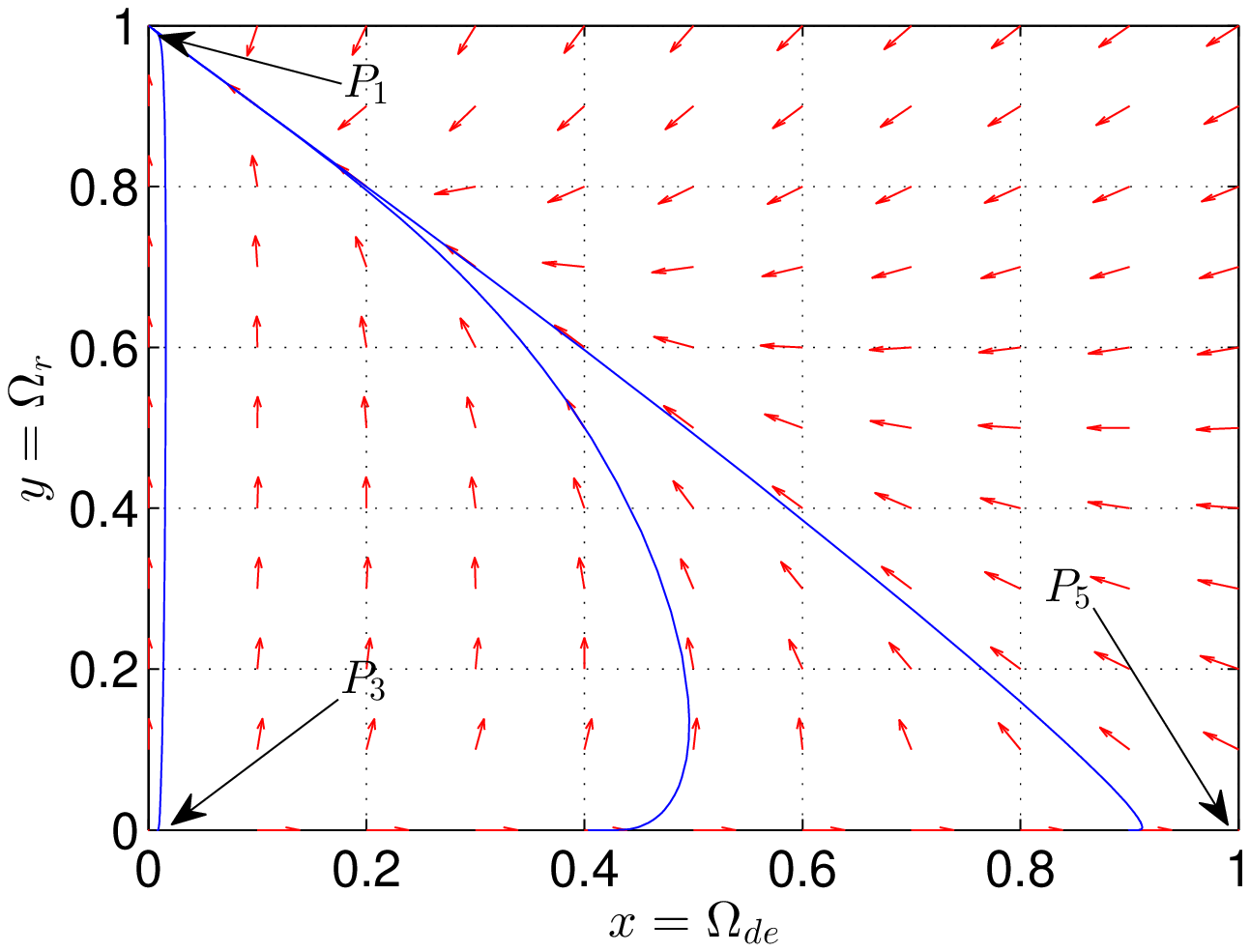}
\caption{\label{figr1}Vector field in plane ($x$, $y$) for the autonomous system (\ref{evx})-(\ref{evy}) with $\gamma_{1}=4/3$ and $\gamma_{2}=1$. The free parameters have been chosen as ($\xi$, $\gamma_{de}$)$=$($4.1$, $0.01$). In this case, the radiation-dominated solution, $P_{1}$ is the late time attractor of the system, representing a viscous phantom solution ($w_{eff}=-1.033$). This latter time attractor solution restricts the system to three critical points ($P_{1}$, $P_{3}$ and $P_{5}$) and none of them corresponds to a true early time RDE or MDE.}
\end{center}
\end{figure}

To be in line with the complete cosmological dynamics, it is necessary to associate an unstable critical point with a RDE. In this model, we can identify two critical points with an unstable behavior, namely $P_{1}$ and $P_{3}$. The first one represents a true RDE with $\Omega_{r}=1$ and $w_{eff}=0$ for $\xi\ll1$, while the latter represents a dark matter solution $\Omega_{m}=1$ and $w_{eff}=0$. The fact that both solutions ($P_{1}$ and $P_{3}$) are unstable simultaneously, when $\gamma_{de}<1$ and $0\leq\xi<1$, implies that we need to choose those initial condition in the neighborhood of $P_{1}$ to guarantee that the RDE is the source of any orbits of cosmological importance in the phase space. Otherwise, it will be impossible to obtain, in terms of $P_{3}$, a successful description of the early time evolution of the Universe. This analysis also implies that those scenarios, under a proper selection of the bulk viscosity parameter, in which $P_{1}$ mimics a MDE ($\xi=1$) or a stable phantom solution ($\xi>4$), are ruled out due to the impossibility of correctly reproducing, in a correct way, the dynamics of the early time Universe. Figure \ref{figr1} shows some example orbits in the phase space to illustrate the situation in which $P_{1}$ mimics a stable phantom solution (late time attractor).  

In addition to a RDE, we also need to associate a saddle critical point with a MDE. The saddle nature of this point guarantees that the cosmological solutions will remain for a lapse of time around it, before ultimately approaching a stable
late time solution. The candidates in this model are $P_{2}$ and $P_{3}$. \footnote{Recall that $P_{1}$ is ruled out as a MDE even if $\xi=1$ ($w_{eff}=0$).} As we mentioned before,  $P_{3}$ always exhibits an unstable nature for a non-null value of the bulk viscosity ($\xi>0$). As a possible past attractor of the model, all the orbits that start at $P_{3}$ describe solutions where there is not a radiation-dominated era preceding this point; thus, $P_{3}$ is ruled out as a true MDE. \footnote{The presence of a RDE is necessary to describe process such as the primordial nucleosynthesis.} Only $P_{2}$ is able to reproduce a true MDE. This critical point represents a scaling solution between radiation and DM fluids and its
existence is a direct consequence of having considered bulk viscosity on the radiation fluid. In order to recover a true MDE ($\Omega_{m}=1$),  the following condition must be met: $\xi \ll 1$. In addition, Tables \ref{tab3} and \ref{tab4} show that, despite it being a scaling solution between radiation and DM, $P_{2}$ is not able to mimic, at background level, a true RDE, even if $\xi\approx1$ ($\Omega_r\approx1$), since $w_{eff}=0$.

\begin{figure}[t!]
\begin{center} 
\includegraphics[width=24cm]{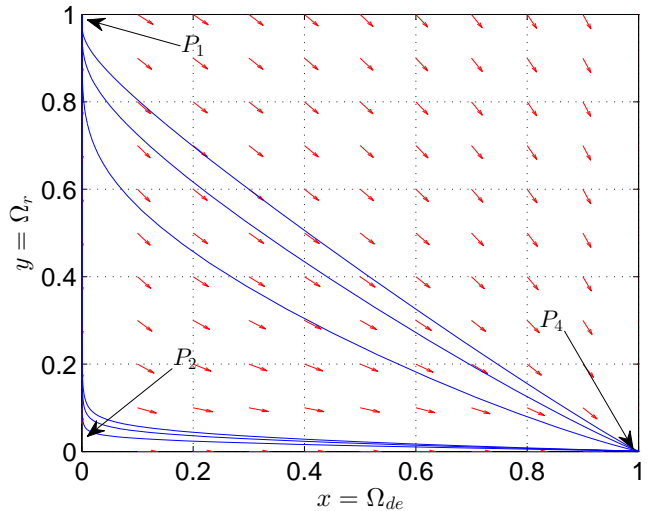}
\caption{\label{figr2}Vector field in the plane ($x$, $y$) for the autonomous system (\ref{evx})-(\ref{evy}) with $\gamma_{1}=4/3$ and $\gamma_{2}=1$. The free parameters have been chosen as ($\xi$, $\gamma_{de}$)$=$($0.01$, $0.01$). In this case, the radiation-DE scaling solution, $P_{4}$, is the late time attractor of the system, representing an accelerated solution. The transition from the RDE ($P_{1}$) to $P_{4}$ allows for the selection of appropriate initial conditions to recover a true MDE ($P_{2}$) with $w_{eff}\simeq0$ and $q\simeq1/2$.}
\end{center}
\end{figure}

Finally, the model has two more accelerated critical points: $P_{4}$ and $P_{5}$.\footnote{Recall that $P_{1}$ is able to mimics an stable phantom solution if $\xi>4$ but this possibility is discarded because the model is not able to describe a RDE and MDE, as shown in Fig. \ref{figr1}.} In the first case, $P_{4}$ represents a stable scaling solution between radiation and dark energy fluids. As shown in Table \ref{tab4}, the presence of a bulk viscosity on the radiation fluid does not induce a crossing of the phantom divide in $P_{4}$, and thus this crossing is only possible if a phantom dark energy fluid is considered in the model ($\gamma_{de}<0$). On the other hand, $P_{5}$ corresponds to a pure dark-energy-domination period, and has saddle behavior for a non-null value of the bulk viscosity. In this model, a favorable scenario would be a transition from $P_{1}$ $\Longrightarrow$ $P_{2}$ $\Longrightarrow$ $P_{4}$: starting in a RDE at early times, then entering into a MDE to finally evolve to a stable accelerated phase. This positive transition is represented in Figure \ref{figr2} through several orbits in the phase space of the model. As we commented before, the existence of a RDE and MDE, in terms of $P_{1}$ and $P_{4}$, respectively, demands that $\xi\ll1$. This latter requirement also implies that the phantom solution with a stable behavior ($P_{1}$ with $\xi>4$) is not compatible with a well-behaved model from the point of view of the complete cosmological dynamics.

\section{Conclusions}

Late time evolutions of a dissipative fluid leading to a universe with phantom behavior have been investigated within different approaches, e.g., \cite{Cataldo2005, Velten2013, Avelino2013}. In this work, our aim was to study the dynamics of a universe filled with radiation, dark matter, and dark energy, where bulk viscosity may be present in one of the matter fluids, with emphasis on the viability of phantom solutions. In the first studied model, the dissipation in the dark matter was characterized by the bulk viscosity proportional to the dark matter energy density, i.e., $\zeta=\zeta_{0}\sqrt{\rho_{m}/\rho_{m0}}$ \cite{Velten2013}; whereas in the second model, the dissipation in the radiation was described by the bulk viscosity proportional to the radiation energy density, i.e., $\zeta=\zeta_{0}\sqrt{\rho_{r}/\rho_{r0}}$ \cite{Velten2013}. In both models, the study was restricted to the cases in which $\zeta\geq0$, a condition that comes from the local second law of thermodynamics \cite{misner1973gravitation, Maartens1996, Zimdahl2000}.

Recall that in  \cite{Avelino2013}, only the case of bulk  viscosity in dark matter was studied. Furthermore, the ansatz on the bulk viscosity now used \cite{Velten2013} is different from that used in \cite{Avelino2013}. Hence, the results obtained in our work are new compared with those obtained in \cite{Avelino2013}.

By making a dynamical system analysis of both possible scenarios and imposing the requirement that both models must follow the so called complete cosmological dynamics, we found that viscous phantom solutions with a stable behavior are not allowed. This results from the fact that it is not possible to recover a viable MDE in the first model if the viscous phantom solution exists ($P_{2}$ with $\xi>3$) and, in the second model, the existence of a stable viscous phantom solution ($P_{1}$ with $\xi>4$) does not allow for the existence of a true RDE and MDE. In other words, there is no smooth transition from a radiation-dominated epoch to a matter-dominated phase to an accelerated late time expansion dominated by dark energy. For the purposes of illustration, we have shown some numerical elaboration for several values of the free parameters of the models ($\xi$, $\gamma_{de}$). 

Additionally, it was shown that, in both models, it is possible to accommodate complete cosmological dynamics whenever $\xi\approx0$. It is noteworthy that these findings corroborate the results obtained in \cite{Velten2012, Velten2013a}. However, in this favorable scenario, the late time attractor solution is characterized by $w_{eff}=-1+\gamma_{de}$; therefore, the nature of this solution depends only on the nature of the dark energy fluid, namely, if the dark energy fluid is a quintessence (phantom) fluid $0<\gamma_{de}<2/3$ ($\gamma_{de}<0$), then the late time stable solution corresponds to a quintessence (phantom) solution. If $\gamma_{de}=0$, then the de Sitter solution will be the late time attractor. 

As we mentioned before, we have focused only on the cases in which $\alpha=1/2$ [see Eq. (\ref{ViscosityDefinition})]. Thus, the question of the viability of the phantom solutions in the Eckart approach, in terms of the complete cosmological dynamical, for $\alpha\neq1/2$ is still an open problem. We hope to address this more general and complex scenario in a forthcoming paper.

\begin{acknowledgments}
    The authors benefited from the many contributions to cosmology in Chile by our recently deceased colleague Sergio del Campo. We acknowledge the support of this research by CONICYT through Grant No. 1140238 (N. C.), Fondecyt Grant No. 1110076 (J. S. and S. L.) and PUCV-VRIEA Grant No. 037.377/2014 (S. L.). Y. L. thanks PUCV for supporting him through Proyecto DI Postdoctorado 2014. F. P. acknowledges Grant No. DI14-0007 of Direcci\'on de Investigaci\'on y Desarrollo, Universidad de La Frontera

\end{acknowledgments}

\bibliography{viscosity}

\end{document}